\documentclass[runningheads]{llncs}
\usepackage{graphicx}
\usepackage{algorithm,algorithmic,times,amsmath,amssymb,latexsym,bm,todonotes,eufrak,array,standalone,lscape,epsfig,enumerate,color,subfigure,graphicx,bussproofs,xcolor,tikz,pdfsync,wrapfig,hyperref,pgfplots}
\usetikzlibrary{shapes.geometric}
\usepackage[all]{xy}
\usepackage[applemac]{inputenc}

\usepackage{colortbl}
\definecolor{myred}{rgb}{0.9,0.1,0.1} 
\definecolor{mygreen}{rgb}{0.1,.7,0.1} 
\definecolor{myblue}{rgb}{0.1,0.1,0.7} 

\begin{document}
\title{A Credal Extension of Independent Choice Logic}
\author{Alessandro Antonucci \and Alessandro Facchini}
\institute{Istituto Dalle Molle di Studi Sull'Intelligenza Artificiale (IDSIA)\\Lugano (Switzerland)\\
\email{\{alessandro,alessandro.facchini\}@idsia.ch}}
\maketitle
\begin{abstract}
We propose an extension of Poole's \emph{independent choice logic} based on a relaxation of the underlying independence assumptions. A \emph{credal} semantics involving multiple joint probability mass functions over the possible worlds is adopted. This represents a conservative approach to probabilistic logic programming achieved by considering \emph{all} the mass functions consistent with the probabilistic facts. This allows to model tasks for which independence among some probabilistic choices cannot be assumed, and a specific dependence model cannot be assessed. Preliminary tests on an object ranking application show that, despite the loose underlying assumptions, informative inferences can be extracted.
\end{abstract}
\section{Introduction}\label{sec:intro}
\emph{Probabilistic logic programming} (PLP) is an emerging area in AI research aiming to develop reasoning tools for relational domains under uncertainty \cite{deraedt15}. PLP can be either intended as a combination of logic programs with probabilistic statements, or a part of the wider current interest in probabilistic programming \cite{carpenter2016stan} specialised to highly structured probability spaces. After some early proposals such as Probabilistic Horn Abduction \cite{poole1993probabilistic} and Independent Choice Logic \cite{poole1997independent}, Distribution Semantics \cite{sato1995statistical} and Probabilistic Datalog \cite{fuhr1995probabilistic}, PLP is currently subject of intense research as well as the theoretical basis for many real-world applications \cite{de2015applications}. 

Most of these theories make assumptions such as the program acyclicity and the mutual independence of the probabilistic facts. This leads to the specification of a single probability mass function over the possible words (\emph{least models}) associated to the logic program. In more recent times, several approaches are trying to go beyond some of these assumptions, but without giving away independence. In some of these cases \cite{cozman2017semantics,luka}, this leads to the adoption of a \emph{credal} semantics, i.e., a PLP does not define a single mass function but a set of joint mass functions over the least models. In this paper we propose a different path consisting in a generalisation of Poole's independent choice logic which keeps the acyclicity condition on programs, but relaxes the independence assumptions. Such an extension can be intended as a credal, conservative, semantics considering \emph{all} the joint mass functions over the least models being also consistent with the probabilistic facts. 
To introduce our motivations for such proposal, consider the following example.

\begin{example}\label{ex:2urns}
{\it Two urns contain red, green and blue balls. Let $\mu_1=[0.60;0.30;0.10]$ and $\mu_2=[0.20;0.35;0.45]$ denote the normalised proportions of the colours in the urns.\footnote{We use semicolons to separate the elements of an array and commas to separate the two bounds of an interval.} Assuming that a ball is randomly drawn from each urn, the joint probability over the nine possible outcomes is the left  matrix in Table~\ref{tab:matrice}. The matrix on the right corresponds to a different situation in which a ball is randomly drawn from the first urn, while the second ball is required to have a colour different from that of the first. To achieve that, an unbiased coin is flipped to decide which one of the two permitted colours should be picked. Both joint mass functions are consistent with the above marginal probabilities (i.e., the sums of the rows/columns are the values in $\mu_1$ and $\mu_2$).}
\end{example}
\begin{table}\label{tab:matrice}
\begin{center}
\begin{tabular}{ccccccc}
\tikz\draw[myred,fill=myred](0,0) circle (.7ex);&
\tikz\draw[mygreen,fill=mygreen](0,0) circle (.7ex);&
\tikz\draw[myblue,fill=myblue](0,0) circle (.7ex);&
$\leftarrow$ ball$_1$ $\rightarrow$&\tikz\draw[myred,fill=myred](0,0) circle (.7ex);&\tikz\draw[mygreen,fill=mygreen](0,0) circle (.7ex);&
\tikz\draw[myblue,fill=myblue](0,0) circle (.7ex);\\
\cellcolor{gray!10}{$0.120$}&\cellcolor{gray!40}{$0.210$}&\cellcolor{gray!40}{$0.270$}&\tikz\draw[myred,fill=myred](0,0) circle (.7ex);& \cellcolor{gray!10}{$0.000$}&\cellcolor{gray!40}{$0.300$}&\cellcolor{gray!40}{$0.300$}\\
\cellcolor{gray!10}{$0.060$}&\cellcolor{gray!10}{$0.105$}&\cellcolor{gray!10}{$0.135$}&\tikz\draw[mygreen,fill=mygreen](0,0) circle (.7ex);&
\cellcolor{gray!10}{$0.150$}&\cellcolor{gray!10}{$0.000$}&\cellcolor{gray!10}{$0.150$}\\
\cellcolor{gray!10}{$0.020$}&\cellcolor{gray!40}{$0.035$}&\cellcolor{gray!40}{$0.045$}
&\tikz\draw[myblue,fill=myblue](0,0) circle (.7ex);&
 \cellcolor{gray!10}{$0.050$}&\cellcolor{gray!40}{$0.050$}&\cellcolor{gray!40}{$0.000$}\\&&&$\uparrow$\\&&&ball$_2$\\
\end{tabular}
\end{center}
\caption{Two joint mass functions over two ternary variables sharing the same marginal probabilities. Dark grey cells corresponds to a query of interest (first ball non-red, second ball non-green).}
\end{table}

This trivial example makes clear that marginal probabilities are in general not sufficient to uniquely determine a joint mass function unless additional assumptions, such as independence or an explicit dependence, are made. If no further assumptions can be made, the most conservative approach corresponds to consider \emph{all} the joint mass functions consistent with the constraints and computing inferences with respect to the multiple probabilistic specification. This is clarified by the following example.

\begin{example}\label{ex:credalurns}
{\it With the same setup as in Example~\ref{ex:2urns}, we want to compute the probability of having the first ball not red and the second not green. Assuming independence, this query has probability $(1-\mu_1(r))(1-\mu_2(g))=0.56$. The same result can be obtained by summing the (consistent) dark grey values in the matrix on the left of Table~\ref{tab:matrice}. Under the coin-based dependence relation, the result becomes instead $0.65$ (sum of the dark grey values in the matrix on the right). With no dependence/independence assumptions, the probability of the query can be only said to belong to the interval $[0.5,0.7]$. These values are the maximum and the minimum obtained from a linear programming task involving the nine joint probabilities as optimisation variables and the consistency with the marginals (together with non-negativity and normalization) as linear constraints.}
\end{example}

The feasible region of the linear programming task in the above example is a convex set of (joint) probability mass functions, i.e., a \emph{credal set} \cite{augustin2014introduction,levi1983enterprise}. Inferences based on credal sets are conservatively intended as the computation of the lower/upper bounds of the query with the mass function varying in the credal set. The interval estimate is also called \emph{imprecise probability}. When coping with relational domains, adding independence statements leading to a unique, ``precise'', specification might be not always a tenable assumption, and the imprecise-probabilistic technique we present in this paper can be regarded as the most conservative approach to the modeling of a condition of \emph{ignorance} about the dependencies among the atoms in a program.

After reviewing some necessary background concepts in Section~\ref{sec:back}, our proposal is first introduced by an example in Section~\ref{sec:motiv} and then formalised as a theory in Section~\ref{sec:ccl}. An inference algorithm is derived in Section~\ref{sec:inference}, while the results of some preliminary experiments are reported in Section~\ref{sec:experiments}. Conclusions and outlooks are in Section~\ref{sec:conc}.
\section{Background}\label{sec:back}
We review here some background information about logic programming and its probabilistic extension.
\paragraph{Logic Programming.} A \emph{term} $t$ is defined as being either a constant or a variable. An \emph{atom} $r(t_1,\ldots,t_k)$ is thence obtained by applying a relational symbol $r$ to a sequence of terms $t_1,\ldots,t_k$. We identify Boolean propositional variables with 0-ary predicates. An atom $a$ or its negation $\lnot a$ is called a \emph{literal}. A \emph{clause} has form
\begin{equation}
a_0 \leftarrow a_1,\ldots,a_m,\neg a_{m+1},\ldots,\neg a_n\,,
\end{equation}
where $a_i$ are atoms for each $i=0,\ldots,n$, $a_0$ is called the \emph{head} of the clause, while the other atoms are the \emph{body}. \emph{Facts} are clauses with empty body. If a clause is not a fact is called a \emph{rule}.  A \emph{ground} atom (literal, clause) is an atom (literal, clause) that does not contain any variable. The \emph{grounding} of a clause is a clause obtained by uniformly replacing constants for the variables in the considered clause. A \emph{logic program} $\mathbf{P}$ is a finite set of clauses. The \emph{Herbrand base} of a program is the set of all ground instances of atoms in the program.  A \emph{query} $Q$ for a program $\mathbf{P}$ is a set of ground literals whose positive part belong to the Herbrand base of $\mathbf{P}$. 
A logic program is \emph{acyclic} if there is an assignment of a positive integer to each element of the Herbrand base such that for every grounding of a rule the number assigned to the head is greater then each number assigned to an element of the body. An \emph{interpretation} $\iota$ of a program $\mathbf{P}$ is a function assigning a truth value to each member of the Herbrand base for $\mathbf{P}$. An interpretation $\iota$ is said to be a stable model for $\mathbf{P}$ if for every ground atom $a$, $a$ is true in $\iota$ if and only if $a$ is a fact in $\mathbf{P}$ or it is the head of the grounding of a rule in $\mathbf{P}$ whose base is true in $\iota$. A negation $\lnot a$ is true in $\iota$ if and only if the atom $a$ is not true in $\iota$. 

The following classical result \cite{apt1991acyclic} is used in the rest of the paper.
\begin{proposition}\label{thm:stable}
An acyclic logic program has a unique stable model.
\end{proposition}

\paragraph{Probabilistic Logic Programming.} There are different ways of representing probabilistic information with logic programming. A possible approach is to annotate or extend clauses with probabilities (see, e.g., \cite{lukasiewicz1998probabilistic,ng1992probabilistic}). Another, nowadays very popular, approach consists in adding to a logic program \emph{independent} probabilistic alternatives representing mutually independent random events with a finite number of different outcomes, such as tossing a coin or rolling a die. Within this latter stream, various languages have been proposed such as  PRISM \cite{sato1995statistical}, ICL \cite{poole1997independent}, pD \cite{fuhr2000probabilistic}, LPAD \cite{vennekens2004logic} or ProbLog \cite{de2007problog}. Despite the differences in their syntactical presentation, under Sato's distribution semantics all these formalisms are in general comparable in their expressive power \cite{lpad}. In this work we focus on Poole's ICL. The choice of this formalism, rather than another among the languages listed above, is due to the fact that its explicit set-theoretical formulation of probabilistic choices will  simplify the presentation of our approach. ICL syntax and semantics are introduced here below.
 
\paragraph{Independent Choice Logic (ICL).} An ICL theory is a triple $\langle\mathbf{P},\mathbf{C},\mu\rangle$ such that:
\begin{itemize}
\item $\mathbf{P}$ is an acyclic logic program.
\item $\mathbf{C}$, called the \emph{choice space}, is a family of non-empty sets of ground atoms. The elements of $\mathbf{C}$ are called \emph{alternatives}, and the elements of each alternative \emph{atomic choices}. Atomic choices from the same or different alternatives cannot unify with each other, nor with the head of any clause in $\mathbf{P}$.
\item $\mu$ specifies a probability mass function over each alternative, i.e., $\mu: \bigcup_{C\in \mathbf{C}} C \to [0,1]$ with $\sum_{a \in C} \mu(a)=1$ for each $C \in \mathbf{C}$.
\end{itemize}

The following example demonstrates the ICL syntax.

\begin{example}\label{ex:icl_syntax}
{\it Consider the setup of Example~\ref{ex:2urns} under the assumption of independence between the colours of the two balls (Table~\ref{tab:matrice} left). Let $a_i^j$ denote the Boolean variable true if and only if the colour $j$ has been drawn from the $i$-th urn. The example corresponds to an ICL theory over the ground atoms $\{a_i^j\}_{i=1,2}^{j=r,g,b}$. The logical program is empty, i.e., $\mathbf{P}=\emptyset$, as the logical constraints over the atoms can be directly embedded in the choice space $\mathbf{C}$. Such a space contains two alternatives, i.e., $\mathbf{C} := \left\{ C_1,C_2 \right\}$, with $C_i=\{a_i^r,a_i^g,a_i^b\}$ for each $i=1,2$. Finally, the probabilities are directly obtained from the marginals, i.e., $\mu(C_i)=\mu_i$ for each $i=1,2$.}
\end{example}

ICL semantics is defined in terms of \emph{possible words}. A \emph{total choice} $\mathfrak{c}$ for the choice space $\mathbf{C}$ is a choice function selecting exactly one atomic choice from each alternative in $\mathbf{C}$, and we associate to $\mathfrak{c}$ a possible world $\omega_\mathfrak{c}$. For each total choice $\mathfrak{c}$ the program $\mathbf{P}\cup\{ a \leftarrow \mid a \in {\sf image}(\mathfrak{c})\}$ is acyclic, and therefore by Proposition~\ref{thm:stable} it has a unique stable model $\iota_\mathfrak{c}$. We thence say that an atom $a$ is true at $\omega_\mathfrak{c}$, written $\omega_\mathfrak{c} \models a$, if $a$ is true in $\iota_\mathfrak{c}$, and states $\omega_\mathfrak{c} \models \lnot a$ if $\omega_\mathfrak{c} \not\models a$.  Given a query $Q$, we write $\omega \models Q$ whenever $\omega \models q$, for any member of $q\in Q$.

As alternatives are assumed to be independent, a ICL theory defines a unique mass function $\mu'$ over the collection $\Omega$ of all the possible worlds as follows:
\begin{equation}\label{eq:ICL_distribution}
\mu'(\omega_\mathfrak{c}):= \prod_{a \in {\sf image}(\mathfrak{c})} \mu(a) \cdot \prod_{a \notin {\sf image}(\mathfrak{c})} (1-\mu(a))\,.
\end{equation}
\begin{example}\label{ex:icl_semantics}
{\it The ICL theory in Example~\ref{ex:icl_syntax} has nine possible worlds corresponding to the elements of $C_1\times C_2$. For each $\omega$, the associated probability assigned by $\mu$ as in Equation~\eqref{eq:ICL_distribution} reproduces the value in Example~\ref{ex:2urns}. The query in Example~\ref{ex:credalurns} is $Q:=\{ \lnot a_1^r, \lnot a_2^g \}$ and it is true only in four possible worlds. The corresponding \emph{success probability} is $\mu'(Q)=\sum_{\omega \models Q}\mu'(\omega) = 0.210+0.270+0.035+0.045$ as in Example~\ref{ex:credalurns}.}
\end{example}
\section{A Motivating Example}\label{sec:motiv}
Consider a possibility space determined by the fact that, on a working day, a person called Andrea is using her car or not, she is working late or not, and that in  Milan, the city where she is living, it is raining or not. We denote by $r$ the atomic fact that it rains in Milan, by $c$ the fact that she uses the car, and by $w$ the fact that she is working late. Hence, the elements of the possibility space we consider correspond to the eight possible worlds associated with $r,c,w$ in Table~\ref{table:poswo}, where $\tt{t}$ denotes true and $\tt{f}$ false.
\begin{table}[htp]
\begin{center}
\begin{tabular}{lc ccccccc}
\hline
& $\omega_1$ & $\omega_2$ & $\omega_3$ &$\omega_4$&$\omega_5$&$\omega_6$&$\omega_7$&$\omega_8$ \\
\hline
$r$ (Rain in Milan)&$\tt{t}$ & $\tt{t}$ & $\tt{t}$ & $\tt{t}$  & $\tt{f}$ & $\tt{f}$ & $\tt{f}$ & $\tt{f}$\\ 
$c$ (Andrea using her car)&$\tt{t}$ & $\tt{t}$ & $\tt{f}$ &  $\tt{f}$ & $\tt{t}$ & $\tt{t}$ & $\tt{f}$ & $\tt{f}$\\ 
$w$ (Andrea working late)&$\tt{t}$ & $\tt{f}$ & $\tt{t}$ &  $\tt{f}$ & $\tt{t}$ & $\tt{f}$ & $\tt{t}$ & $\tt{f}$\\
\hline
$\mu'(\omega_i)$&0.01&0.04&0.01&0.04&0.09&0.36&0.09&0.36\\
\hline
$\mu^{(1)}(\omega_i)$&0.1&0.0&0.0&0.0&0.0&0.4&0.1&0.4\\
$\mu^{(2)}(\omega_i)$&0.1&0.0&0.0&0.0&0.1&0.3&0.0&0.5\\
$\mu^{(3)}(\omega_i)$&0.0&0.1&0.0&0.0&0.0&0.4&0.2&0.3\\
$\mu^{(4)}(\omega_i)$&0.0&0.1&0.0&0.0&0.2&0.2&0.0&0.5\\
$\mu^{(5)}(\omega_i)$&0.0&0.0&0.1&0.0&0.1&0.4&0.0&0.4\\
$\mu^{(6)}(\omega_i)$&0.0&0.0&0.1&0.0&0.0&0.5&0.1&0.3\\
$\mu^{(7)}(\omega_i)$&0.0&0.0&0.0&0.1&0.2&0.3&0.0&0.4\\
$\mu^{(8)}(\omega_i)$&0.0&0.0&0.0&0.1&0.0&0.5&0.2&0.2\\
\hline
\end{tabular}
\end{center}
\caption{A possibility space as a set of possible worlds, and nine mass functions over it.}
\label{table:poswo}
\end{table}

Let us compute the probability that Andrea will hang out with friends under the following assumptions.
\begin{itemize}
\item The probability of rain in Milan is $0.1$, that for  Andrea using her car $0.5$ and that for her working late $0.2$.
\item If Andrea is with her car or it is raining, she will visit her parents.
\item If she is neither visiting her parents nor working late, then she will be hanging out with friends.
\end{itemize}
Let $p$ denote the fact that Andrea visits her parents, and $h$ the fact associated to her hanging out with friends. In a PLP dialect such as ICL, if we introduce symbols $nr$, $nc$ and $nw$ to denote the complementary atomic facts associated respectively to $r$, $c$ and $w$, the example corresponds to a theory with:
\begin{itemize}
\item $\bm{C}=\{C_1,C_2,C_3\}$, with alternatives $C_1=\{ r, nr\}$, $C_2=\{ c, nc\}$, $C_3=\{ w, nw\}$;
\item $\mu(r)=0.1$, $\mu(nr)=0.9$, $\mu(c)=0.5$, $\mu(nc)=0.5$, $\mu(w)=0.2$, $\mu(nw)=0.8$;
\item and ${\bf P}=\{p \leftarrow c, p \leftarrow r, h \leftarrow \lnot p, nw \}$. 
 \end{itemize}
Hence, for instance, $\omega_i \models c$ if and only if $\omega_i \not\models nc$ if and only if $\omega_i \not\models \lnot c$, with $i\in \{1, \dots, 8 \}$. It also holds that $\omega_i \models p$, for $i\in \{1, \dots, 6 \}$ and $\omega_i \models \lnot p$, for $i=7,8$. Similarly $\omega_8 \models h$ and $\omega_i \models \lnot h$ for $i \neq 8$. Since alternatives are assumed to be independent, a unique probability mass function $\mu'$ (see Table~\ref{table:poswo}) is defined over $\Omega:=\{\omega_1, \dots, \omega_8\}$ and, for instance, the success probability of the query $h$ is $\mu'(\omega_8):=0.9\cdot 0.5\cdot 0.8 = 0.36$. Yet, it is easy to identify realistic situations in which such independence assumptions are violated. E.g., Andrea using her car might be not independent of raining in Milan. While the modelling of a deterministic dependence (e.g., $c \leftarrow r$) or a probabilistic influence (e.g., $\mu(c\mid r)=0.95$ and $\mu(c\mid nr)=0.45$) can be described in standard PLPs, a condition of complete ignorance about the relations between two or more atomic facts requires a generalisation of the semantics. For instance, if we do not make any assumption about the (in)dependence relations among the possible choices corresponding to the three variables in our example, the whole set of mass functions consistent with the marginals is the convex hull of the eight mass functions $\{ \mu^{(i)} \}_{i=1}^8$ in Table~\ref{table:poswo} and the probability of the query $h$ can only be said to belong to the range $[0.2,0.5]$. This interval shrinks to $[0.32,0.40]$ if no assumptions can be made about the independence between choices on elements of $C_1$ and choices on elements of $C_2$, but both are assumed to be independent with respect to choices on $C_3$.

In the next section we formalise these ideas in a general framework, called \emph{credal choice logic} (CCL), which provides an extension of Poole's independent choice logic in which the independence condition is relaxed.
\section{Credal Choice Logic}\label{sec:ccl}
\paragraph{Syntax.} From a syntactical point of view, the idea pursued in this paper is to consider elements of the choice space not as independent alternatives but as \emph{independent families}, each including possibly correlated alternatives. This is formally stated in the following definition.
\begin{definition}
A CCL theory $\mathcal{T}$ is a triple $\langle \mathbf{P}, \mathcal{C}, \mu \rangle$ where:
\begin{itemize}
\item $\mathbf{P}$ is an acyclic logic program.
\item $\mathcal{C}=\{\mathbf{C}_1, \dots, \mathbf{C}_k\}$ is a family of choice spaces (i.e., a set of sets of non-empty sets of ground atoms).  Alternatives and atomic choices are intended as in ICL. In particular we assume that:
\begin{enumerate}
\item for each choice space $\mathbf{C} \in \mathcal{C}$, no atomic choice in $\bigcup \mathbf{C}$ should unify with the head of any clause in $\mathbf{P}$; 
\item for each pair $\mathbf{C}, \mathbf{C}' \in \mathcal{C}$ with $\mathbf{C}\neq \mathbf{C}'$, the sets of atomic choices $\bigcup \mathbf{C}$ and  $\bigcup \mathbf{C}'$ are disjoint.
\end{enumerate}
\item $\mu$ specifies a probability mass function over each alternative of each choice space, i.e., $\mu: \bigcup \bigcup \mathcal{C} \to [0,1]$ with $\sum_{a \in C} \mu(a)=1$, for every $C \in \mathbf{C}$, and every $\mathbf{C} \in \mathcal{C}$.
\end{itemize}
\end{definition}
Contrary to ICL, in a CCL theory the alternatives in a given choice space are not assumed  to be independent. Consequently there could be a choice space $\mathbf{C}$ with alternatives $C_1,C_2\in\mathbf{C}$ such that $C_1\neq C_2$ but $C_1 \cap C_1 \neq \emptyset$. On the other hand, any ICL theory can be formalised as a CCL theory $\langle\mathbf{P},\mathcal{C},\mu\rangle$ whose choice spaces $\mathbf{C} \in \mathcal{C}$ are singletons.
\begin{example}\label{ex:friends}
{\it Dropping the independence assumption about the relation between Andrea going by car and raining in Milan in the example in the previous section transforms the corresponding ICL theory in a CCL theory $\mathcal{T}_{\sf friends}:=\langle {\bf P}, \{\mathbf{C}_1, \mathbf{C}_2\}, \mu \rangle$, where 
${\bf C}_1= \{ C_1, C_2\}$, ${\bf C}_2= \{ C_3\}$, while ${\bf P}$ and $\mu$ are defined as in the original example. The fact fact that alternatives $C_1$ and $C_2$ belong to the same choice space means that the choice among elements of $C_1$ (namely between $c$ and $nc$) is not assumed to be independent from the choice among elements of $C_2$ (namely between $r$ and $nr$). On the other hand the fact that the alternative $C_3$ does not belong to the same choice space as $C_1$ and $C_2$ means that the choice between $w$ and $nw$  is independent of the two previous choices.}
\end{example}
\paragraph{Semantics.} As in ICL, the CCL semantics is defined in terms of possible words. A total choice for a family $\mathcal{C}$ of choice spaces is a choice function $\mathfrak{c}_i$ on $\bigcup\mathcal{C}$ such that, for $C$ and $C'$ belonging to the same choice space $\mathbf{C}$, the following coherence condition is satisfied:
\begin{equation}\label{eq:coherence}
\mathfrak{c}_i(C)=\mathfrak{c}_i(C')\,,
\end{equation}
whenever $\mathfrak{c}_i(C) \in C \cap C'$. Hence a total choice selects coherently (in the sense of Equation~\eqref{eq:coherence}) exactly one atomic choice from each alternative of every choice space of $\mathcal{C}$.

Now, consider the program $\mathbf{P}(\mathfrak{c}):=\mathbf{P} \cup \{ a \leftarrow \; \mid \varphi \in {\sf image}(\mathfrak{c})\}$. From the disjointness conditions, $\mathbf{P}(\mathfrak{c})$ is necessarily acyclic and therefore by Proposition~\ref{thm:stable} it has a unique stable model $I_{(\mathfrak{c})}$. As for ICL, for each total choice $\mathfrak{c}$ we define a corresponding possible world $\omega_\mathfrak{c}$ and the associated notion of being true in it. 

In the case of ICL, the probability for a possible world is given as in Equation~\eqref{eq:ICL_distribution} by the product of the probabilities of the atomic choices true in it and one minus the probabilities of the atomic choices false in it. The distribution semantics we defined for CCL acts similarly but on sets of probabilities. 

To illustrate the idea, notice that a total choice $\mathfrak{c}$ restricted to a choice space $\mathbf{C}_i$ determines a partial selection $\mathfrak{c}_i$. Such partial selection can be identified with all total choices extending it, and thus with the collection $E_{\mathfrak{c}_i}$ of possible world in which all elements in the image of $\mathfrak{c}_i$ are true. Such collection represents the possible worlds an agent cannot tell apart if she is given only the partial information provided by $\mathfrak{c}_i$. Let  $\Omega_i$ be the set of all collections associated with $\mathbf{C}_i$. Assume $\mu_i$ is a probability mass function over $\Omega_i$, and define $\mu_i(a):=\sum_{\mathfrak{c}_i: {a \in {\sf image}(\mathfrak{c}_i)}} \mu_i(E_{\mathfrak{c}_i})$, for $a \in \bigcup \mathbf{C}_i$.  We say that $\mu_i$ \emph{agrees} with $\mu$ on $\mathbf{C}_i$ if $\mu_i(a)=\mu(a)$, for every $a \in \bigcup\mathbf{C}_i$. Hence, call $\mathcal{M}_i$ the set of all probability distributions over $\Omega_i$ agreeing with $\mu$ on $\mathbf{C}_i$.

Each possible world $\omega_\mathfrak{c}\in \Omega$ can be identified with the unique sequence $(E_{\mathfrak{c}_i}: i\in\{1,\dots, k\}) \in \times_{i\in\{1,\dots, k\}}\Omega_i$ such that $\{\omega_\mathfrak{c}\}=\bigcap_{i\in\{1,\dots, k\}}E_{\mathfrak{c}_i}$. For $\mu_i \in \mathcal{M}_i$ with $ i\in\{1,\dots, k\}$, let $\mu':= \prod_{i\in\{1,\dots, k\}} \mu_i$ be the mass function on $\Omega$ defined by:	
\begin{equation}\label{eq:facto1}
\mu'(\omega_\mathfrak{c})=\prod_{i\in\{1,\dots, k\}} \mu_i(E_{\mathfrak{c}_i})\,.
\end{equation}
We can therefore associate with a CCL theory $\mathcal{T}$ a joint credal set $\mathcal{M}_\mathcal{T}$ obtained as the following set of factorising mass functions:
\begin{equation}\label{eq:credal}
\left\{
\prod_{i_{i\in\{1,\dots, k\}}} \mu_i 
\mid 
\mu_i \in \mathcal{M}_i, i\in\{1,\dots, k\}
\right\}\,.
\end{equation}
It is immediate to verify the following result.
\begin{proposition}
Let $\mathcal{T}$ be a CCL theory. $\mathcal{M}_\mathcal{T}$ is the largest closed convex set of probability mass functions $\mu'$ on $\Omega$ that agree with $\mu$, that is $\mu' (a)=\mu(a)$ for every $a \in \bigcup\bigcup \mathcal{C}$.
\end{proposition}
In the imprecise-probability jargon, the condition in Equation~\ref{eq:credal} is called \emph{strong independence} between the variables
associated to the different choice spaces \cite{augustin2014introduction}. Accordingly, we call the credal set $\mathcal{M}_\mathcal{T}$ \emph{strong extension} of the marginals specified by $\mu$. In particular, we have that for $\mu_i \in \mathcal{M}_i$ there is $\mu' \in  \mathcal{M}_\mathcal{T}$ such that:
\begin{equation}\label{eq:facto2}
\mu_i(E_{\mathfrak{c}_i})= \sum_{	\omega \in E_{\mathfrak{c}_i}}\mu'(\omega)=\mu'({\sf image}(\mathfrak{c}_i))\,.
\end{equation} 
Given a CCL theory $\mathcal{T}$, a question is therefore whether or not a probability mass function over the possible worlds consistent with the marginals exists. By Equation~\eqref{eq:credal} it is enough to answer such question when $\mathcal{C}$ contains a single choice space. But this can be shown by induction on the number of alternatives. Hence, the following holds.
\begin{proposition}\label{pr:nonempty}
The credal set $\mathcal{M}_\mathcal{T}$ of a CCL theory $\mathcal{T}$ is non-empty.
\end{proposition}
\begin{example}\label{ex:friends2}
{\it  Consider the CCL theory $\mathcal{T}_{\sf friends}:=\langle {\bf P}, \{\mathbf{C}_1, \mathbf{C}_2\}, \mu \rangle$ in Example~\ref{ex:friends}. There are four partial choices $\mathfrak{c}^j_1$ with respect to $\mathbf{C}_1$ and thus four collections $E_{\mathfrak{c}^j_1}$ of possible worlds and, similarly, two partial choices $\mathfrak{c}^j_2$ with respect to $\mathbf{C}_2$ and two collections $E_{\mathfrak{c}^j_1}$ of possible worlds (see Table \ref{table:poswo2}). We have that $\mathcal{M}_1$ is the collection of all probability mass functions  $\mu_1: \{E_{\mathfrak{c}^1_1}, \dots, E_{\mathfrak{c}^4_1} \} \to [0,1]$ such that $\mu(r)=\mu_1(E_{\mathfrak{c}^1_1})+\mu_1(E_{\mathfrak{c}^2_1})$, $\mu(nr) = \mu_1(E_{\mathfrak{c}^3_1})+\mu_1(E_{\mathfrak{c}^4_1})$, $\mu(c)=\mu_1(E_{\mathfrak{c}^1_1})+\mu_1(E_{\mathfrak{c}^3_1})$ and $\mu(nc)= \mu_1(E_{\mathfrak{c}^2_1})+\mu_1(E_{\mathfrak{c}^4_1})$. On the other hand, $\mathcal{M}_2$ is given by taking $\mu_2: \{E_{\mathfrak{c}^1_2},  E_{\mathfrak{c}^2_2} \} \to [0,1]$ defined simply as $\mu_2(E_{\mathfrak{c}^1_2}) = \mu(w)$ and $\mu_2(E_{\mathfrak{c}^2_2}) = \mu(nw)$. Hence $\mathcal{M}_{\mathcal{T}_{\sf friends}}= \{ \mu'= \mu_1\cdot\mu_2 \mid \mu_1 \in \mathcal{M}_1,\mu_2 \in \mathcal{M}_2\}$, where $\mathcal{M}_1$ is the convex hull of $\mu_1^{(1)} = [0.0; 0.1;0.4;0.5]$ and $\mu_1^{(2)} = [0.1;0.0;      0.5;0.4]$, while $\mathcal{M}_2$ has a single element $\mu_2 = [0.2;0.8]$.} 
\begin{table}[htp!]
\begin{center}
\begin{tabular}{lcccccc}
\hline
{$(i,j)$}&{$(1,1)$}&{$(1,2)$}&{$(1,3)$}&{$(1,4)$}&{$(2,1)$}&{$(2,2)$}\\
\hline
$E_{\mathfrak{c}_i^j}$&
$\{\omega_1,\omega_2\}$&$\{\omega_3,\omega_4\}$&$\{\omega_5,\omega_6\}$&$\{\omega_7,\omega_8\}$ &
$\{\omega_1,\omega_3,\omega_5,\omega_7\}$ & $\{\omega_2,\omega_4,\omega_6,\omega_8\}$ \\
${\sf image}(\mathfrak{c}_i^j)$ &$\{r,c\}$  & $\{r,nc\}$  & $\{nr,c\}$  & $\{nr,nc\}$ & $\{w\}$ & $\{nw\}$\\ 
\hline
\end{tabular}
\caption{Classes (of possible worlds) for $\mathcal{M}_1$ and $\mathcal{M}_2$.}
\label{table:poswo2}
\end{center}
\end{table}
\end{example}

\section{Inference}\label{sec:inference}
In the previous section, by relaxing the independence assumptions, we introduced CCL as a  generalisation of ICL. Here we discuss how to compute inferences in CCL. 

Given a CCL theory $\mathcal{T}$, the CCL semantics leads to the specification of a credal set $\mathcal{M}_{\mathcal{T}}$ over $\Omega$ called the \emph{strong extension} of $\mathcal{T}$. Given a \emph{query} $Q$ (see definition in Section~\ref{sec:back}) for the program $\mathbf{P}$ in $\mathcal{T}$, inference is intended as the computation of the lower $\underline{\mu}(Q)$ and upper $\overline{\mu}(Q)$ bounds of the success probability of the query with respect to the strong extension, i.e.,
\begin{equation}\label{eq:inf}
\underline{\mu}(Q):=\min_{\mu(\Omega) \in \mathcal{M}_\mathcal{T}} \sum_{\omega \models Q} \mu(\omega)\,,
\end{equation}
and similarly, with the maximum replacing the minimum for the upper success probability $\overline{\mu}(Q)$. As $\mathcal{M}_\mathcal{T}$ is non-empty (Proposition~\ref{pr:nonempty}), Equation~\eqref{eq:inf} always returns proper inferences. Moreover, being induced by a finite number of linear constraints over the joint probabilities (non-negativity, normalisation, and consistency with the marginal probabilistic facts), the convex set $\mathcal{M}_{\mathcal{T}}$ has only a finite number of extreme points and Equation~\eqref{eq:inf} can be regarded as a linear programming task, whose solution corresponds to an extreme point of $\mathcal{M}_\mathcal{T}$. E.g., the interval $[0.2,0.5]$ obtained in Section~\ref{sec:motiv} for the query $h$ under a complete relaxation of the independence assumptions can be regarded as a CCL inference over a strong extension corresponding to the convex hull of the mass functions $\{\mu^{(i)}\}_{i=1}^8$ in Table~\ref{table:poswo}. Such a brute-force approach cannot be used in general as the number of extreme points might be exponentially large with respect to the input size. In the rest of this section we show that even if, as expected, inference in CCL is NP-hard, these inferences can be reduced to a classical task for which dedicated solver and algorithms have been developed.

To do that, let us first denote as CCL$_\alpha$ the decision task associated to an inference, i.e., given a $\alpha\in[0,1]$, deciding whether or not $\alpha \in [\underline{\mu}(Q),\overline{\mu}(Q)]$. In case the theory contains a single choice space, CCL$_\alpha$ can be reduced to \emph{probabilistic satisfiability} (PSAT).  PSAT is a generalisation of the classical \emph{satisfiability} (SAT) decision problem, in which each element of a finite set of Boolean formulas is paired with a probabilistic assessment \cite{nilsson}. The task is to decide whether or not the assessments are consistent, i.e., to determine whether or not a joint mass function over the variable assigning the given probability to each formula exists. A PSAT instance is thus a set $\{ P(\varphi_i) = \alpha_i \mid i< n\}$ where each $\varphi_i$ is a Boolean formula and $\alpha_i$ is the associated probability. As the problem generalises SAT for propositional calculus it is NP-hard. The NP-completeness of the task has been proved in \cite{georgakopoulos} and, unlike SAT, PSAT remains NP-complete even if each clause does not contains more than two literals. 

We first describe how PSAT can be used to solve CCL$_\alpha$. Let $\langle \mathbf{P}, \mathcal{C}=\{ \mathbf{C}\}, \mu \rangle$ be a CCL theory with a single choice space. First, we convert $\mathbf{P}$ to an equivalent Boolean formula $\varphi_\mathbf{P}$ (see, e.g., \cite{janhunen2004representing} for some conversions), and encode the choice space $\mathbf{C}$ as a conjunction of XOR:
\begin{equation}
\varphi_\mathbf{C}:= \bigwedge_{C \in \mathbf{C}} \bigoplus_{a \in C} a\,.
\end{equation}
The class of models of $\varphi_ \mathbf{C}\land \varphi_ \mathbf{P}$ corresponds to the stable models of the original CCL theory. The associated PSAT instance is therefore the set $\Sigma_{\alpha}:= \{ P(\varphi_ \mathbf{C}\land \varphi_ \mathbf{P}) = 1\} \cup \{ P(a)= \mu(a) \mid a \in \bigcup  \mathbf{C}\} \cup \{P(\bigwedge Q)=\alpha\}$.  Clearly $\alpha \in [\underline{\mu}(Q),\overline{\mu}(Q)]$ if and only if $\Sigma_{\alpha}$ is satisfiable. With a small abuse of notation, we denote as SAT(CCL$_\alpha$) a Boolean variable true if and only if $\Sigma_{\alpha}$ is satisfiable.

\begin{figure}[htp!]
\begin{center}
\begin{tikzpicture}[thick]
\draw [black] (-.5,0) -- (5.5,0);
\node at (0,.4) {0};
\node at (5,.4) {1};
\draw [fill,blue] (2.75,0) circle [radius=0.05];
\draw [fill,blue] (1.2,0) circle [radius=0.05];
\draw [fill,black] (2,0) circle [radius=0.05];
\draw [fill,red] (3.5,0) circle [radius=0.05];
\draw [fill,red] (.5,0) circle [radius=0.05];
\node [black] at (1,.4) {$\underline{\mu}(Q)$};
\node [black] at (3,.4) {$\overline{\mu}(Q)$};
\node [black] at (2,.4) {$\mu_I(Q)$};
\node [red] at (3.5,-.4) {\tiny UNSAT};
\node [red] at (.5,-.4) {\tiny UNSAT};
\node [blue] at (1.2,-.4) {\tiny SAT};
\node [blue] at (2.75,-.4) {\tiny SAT};
\node [black] at (2,-.4) {\tiny SAT};
\draw [black] (3,.1) -- (3,-.1);
\draw [black] (1,.1) -- (1,-.1);
\draw [black] (0,.2) -- (0,-.2);
\draw [black] (5,.2) -- (5,-.2);
\end{tikzpicture}
\end{center}
\caption{Inner and outer approximations of CCL inferences.}
\label{fig:bracketing}
\end{figure}
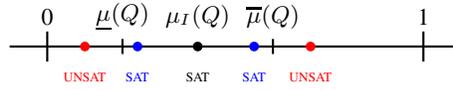

The computation of $[\underline{\mu}(Q),\overline{\mu}(Q)]$ can be therefore achieved by iterating the above reduction for different values of $\alpha$ according to a bracketing scheme that identifies both the bounds (Figure~\ref{fig:bracketing}). The procedure is basically a bisection method that recognises UNSAT values of $\alpha$ as outer approximations (red points in Figure~\ref{fig:bracketing}) and SAT values as inner approximations (blue points in Figure~\ref{fig:bracketing}). To solve PSAT instances, solvers such of those developed by \cite{finger11} and \cite{cozman2015probabilistic} can be used. To achieve a precision $\epsilon$ in the estimates, the number of calls of these solvers is $O(\log \epsilon^{-1})$.


A key point for the bracketing algorithm is the computation of an inner (i.e., SAT) value $\mu_I(Q)$ (e.g., black point in Figure~\ref{fig:bracketing}). We achieve this by translating it into the inference task of computing  in ProbLog  the marginal probability of the query given some specific evidence. More precisely, we first translate $\mathcal{T}$ into ProbLog by distinguishing different occurrences of the same atomic choices and adding, for each duplicated atomic choice, a clause of the form $f \leftarrow a,a'$, where $f$ is some fixed new symbol. Reading $f$ as denoting the false, the newly introduced clauses mimics the coherence conditions on total choices defined on CCL theories. Then, we modify the probabilistic assignment $\mu$ to each atomic choice $a$ in such a way that the marginal probability of $a$ given the evidence $f= {\tt f}$ (i.e. $f$ is false) coincide with $\mu(a)$. Hence, the value $\mu_I(Q)$ corresponds  to the ProbLog computation of the marginal probability for the query $Q$ given the evidence $f={\tt f}$. Such task can be reduced to weighted model counting \cite{fierens2015inference}.


The above procedure should be adopted when coping with CCL theories having a single choice space as in the case of object ranking (see Section~\ref{sec:experiments}). For the general case, we suggest the following (outer) approximation scheme:
\begin{equation}\label{eq:inf2}
\underline{\mu}(Q) \geq \sum_{\omega_{\mathfrak{c}} \models Q} \underline{\mu}(\omega_{\mathfrak{c}}) =  \sum_{\omega_{\mathfrak{c}} \models Q} \prod_{i=1}^k \underline{\mu}({\sf image}(\mathfrak{c}_i))\,,
\end{equation}
where the inequality follows from Equation~\eqref{eq:inf} by simply swapping the minimum and the sum, while the equality with the term on the right-hand side follows from the factorisation in Equation~\eqref{eq:facto1} and then by applying Equation~\eqref{eq:facto2}. 

\begin{example}
{\it Consider the CCL theory $\mathcal{T}_{\sf friends}$ from Example~\ref{ex:friends}. We calculate the  lower success probability of query $h$.
As the theory contains multiple choice spaces, we should use Equation~\eqref{eq:inf2}. However notice that $\omega_i \models h$ exactly when $i=8$ (see Section~\ref{sec:motiv}) and therefore:
\begin{equation*}
\underline{\mu}(\{h\})=\underline{\mu}(\omega_8)=\underline{\mu}_1(\{nr,nc\}) \cdot \underline{\mu}_2(\{nw\})\,,
\end{equation*}
which gives the numerical value $0.32$, as expected from Section~\ref{sec:motiv}. Analogously we get $0.40$ for the corresponding upper success probability.}
\end{example}

\section{Empirical Analysis}\label{sec:experiments}
In this section we report the results of a very first empirical analysis of the approach to CCL inference described in the previous section. Unlike most of the statistical models based on credal sets \cite{augustin2014introduction}, CCL has no parameters directly affecting the \emph{imprecision} level of the inferences (i.e., the difference between the upper and the lower probability of a query). A first important question is therefore whether or not, our relaxation of the standard independence assumptions in PLP, allows to obtain non-vacuous inferences from a query. Another point is whether or not available PSAT algorithms are able to solve instances induced by CCL theories.\footnote{Here we use the solver  \cite{finger11}, freely available at \url{http://psat.sourceforge.net}.} 

\begin{figure}[htp]
\begin{center}
\begin{tikzpicture}[thick]
\node at (0,0) {\tiny $a\succ b\succ c \times 3$ };
\node at (0,.4) {\tiny $a\succ c\succ b \times 5$};
\node at (0,.8) {\tiny $b\succ a\succ c \times 2$};
\node at (0,1.2) {\tiny $b\succ c\succ a \times 4$};
\node at (0,1.6) {\tiny $c\succ a\succ b \times 3$};
\node at (0,2) {\tiny $c\succ b\succ a \times 1$};
\draw[->, >=latex,gray,line width=5pt](1,1) to (3,1) ;
\node at (4,1.8) {\tiny $a$};
\node at (4.4,1.8) {\tiny $b$};
\node at (4.8,1.8) {\tiny $c$};
\node at (3.4,1.4) {\tiny first};
\node at (3.4,1.0) {\tiny second};
\node at (3.4,.6) {\tiny third};
\node at (4,1) [minimum size=.4cm,draw] (a) {\tiny $5$};
\node at (4.4,1) [minimum size=.4cm,draw] (a) {\tiny $4$};
\node at (4.8,1) [minimum size=.4cm,draw] (a) {\tiny $9$};
\node at (4,1.4) [minimum size=.4cm,draw] (a) {\tiny $8$};
\node at (4.4,1.4) [minimum size=.4cm,draw] (a) {\tiny $6$};
\node at (4.8,1.4) [minimum size=.4cm,draw] (a) {\tiny $4$};
\node at (4,0.6) [minimum size=.4cm,draw] (a) {\tiny $5$};
\node at (4.4,0.6) [minimum size=.4cm,draw] (a) {\tiny $8$};
\node at (4.8,0.6) [minimum size=.4cm,draw] (a) {\tiny $5$};
\end{tikzpicture}
\end{center}
\caption{From complete rankings to marginal counts.}
\label{fig:rank2counts}
\end{figure}
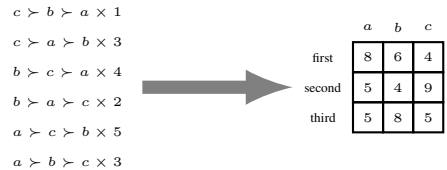

To achieve that we consider \emph{object ranking}, that is the task of deriving a complete ranking over a set of $n$ objects from a data set $\mathcal{D}$ of complete rankings (if features are also considered, the term preference/label learning is used instead) \cite{eyke}. We assume rankings not directly available: only the $n^2$ \emph{marginal} counts reporting how many times a certain object gets a certain rank are available. Note that, in general, complete information about the rankings cannot be recovered from these counts (Figure~\ref{fig:rank2counts}). 

Consider for instance a horse race. Call $r_j$ the unary predicate denoting the property ``ending the race in $j$-th position'', and $h_i$ the constant associated to the $i$-th horse with $i,j=1,\ldots,n$. The ground atom $r_j(h_i)$ means ``horse $i$  ended the race in $j$-th position''. Under our assumptions, we can learn from the data the (marginal) probabilities $\alpha_{i,j}:=\mu(r_j(h_i))$. Object ranking can be based on a joint mass function over these $n^2$ ground toms. This function should reproduce the marginal probabilities being also consistent with the obvious logical constraints (``one and only one horse ends the race in $j$-th position'', and ``one and only one position is reached by the $i$-th horse at the end of the race''). These constraints are encoded in the following formula: 
\begin{equation}\label{eq:constr}
\varphi_{\sf ranking}:=\bigwedge_{j=1}^n \bigoplus_{i=1}^n r_j(h_i) \land \bigwedge_{i=1}^n\bigoplus_{j=1}^n r_j(h_i)\,,
\end{equation}
where $\oplus$ is the exclusive disjunction (XOR). Algebraically, the task corresponds therefore to the specification of a joint mass function over $n^2$ Boolean variables, i.e., $2^{n^2}$ probabilities. Only $n!$ probabilities are non-zero as soon as we impose the 
constraints in Equation~\eqref{eq:constr}, while the $n^2$ marginal probability induce an equal number of (linear) constraints. Thus, we might have non-unique 
specifications even with four objects only.

Consistently with what above, although complete rankings over $n$ objects are available for training, we only learn the $n^2$ marginal probabilities. This is done  by smoothing the frequencies with a Laplace prior of equivalent size two. The corresponding CCL theory is defined as the program-free CCL theory $\mathcal{T}_{\sf ranking}:=\langle \emptyset, \mathcal{C}, \mu \rangle$ with a single choice space, i.e., $\mathcal{C}= \{ \{ C_1, \dots, C_n\} \cup \{ C'_1, \dots, C'_n\}\}$  with $C_\ell=\{r_{1}(h_\ell),\dots,r_{n}(h_\ell)\}$ and $C'_\ell = \{r_{\ell}(h_1),\dots,r_{\ell}(h_n)\}$, and by stating $\mu(r_j(h_i)):=\alpha_{i,j}$ for each $i,j=1,\ldots,n$. This eventually induces PSAT instances with the same probabilistic facts and logical constraints obtained by a CNF (\emph{conjunctive normal form}) conversion of Equation~\eqref{eq:constr}.\footnote{A XOR rewrites as a disjunction together with negations of pairwise conjunctions.}



As a first benchmark, we use four classical UCI datasets (Vehicle, Stock, Glasses, and Bodyfat) with $n=4,5,6,7$. We evaluate pairwise preferences between pairs of objects. We write $i'\succ i''$ to denote the fact that object $i'$ has a higher ranking than object $i''$. To do that, we extend the theory $\mathcal{T}_{\sf ranking}$ by adding to the (initially empty) program $\mathbf{P}$, the following clauses:
\begin{equation}\label{eq:pairs}
\left\{q \leftarrow r_{j'}(h_{i'}), r_{j''}(h_{i''}) \mid \substack{j',j''=1,\dots, n\\ s.t. \quad j'>j''} \right\}\,,
\end{equation}
where $q$ is a new symbol. Hence, the query corresponding to the property $i'\succ i''$ will be given by $Q_{i'\succ i''}:=\{q\}$. For this query we compute both $\underline{\mu}(Q_{i'\succ i''})$ and $\overline{\mu}(Q_{i'\succ i''})$ and, on the basis of these values, we decide whether $i'\succ i''$ or $i''\succ i'$. Note that when coping with imprecise, interval-valued, inferences, a condition of indecision between the two options can be also observed.

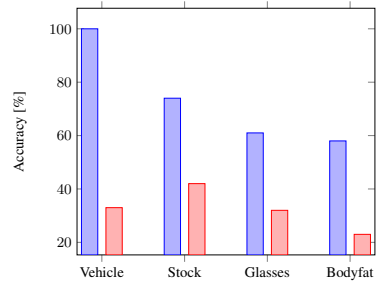
\begin{figure}[htp!]
\begin{center}
\begin{tikzpicture}[scale=.617]
\begin{axis}[
xlabel={},
ylabel={Accuracy [\%]},
width=8cm,
ybar=5pt,
xtick=data,
symbolic x coords={Vehicle,Stock,Glasses,Bodyfat}]
\addplot coordinates {(Vehicle,100) (Stock,74) (Glasses,61) (Bodyfat,58)};
\addplot coordinates {(Vehicle,33) (Stock,42) (Glasses,32) (Bodyfat,23)};
\end{axis}
\end{tikzpicture}
\caption{ICL accuracy for preference 
queries on instances in which CCL is determinate (blue) and indeterminate (red).\label{fig:exps}}
\end{center}
\end{figure}

This is the case if 
$[\underline{\mu}(Q_{i'\succ i''}),\overline{\mu}(Q_{i'\succ i''})]$ overlaps the decision threshold $.5$, otherwise a clear preference is returned. This decision (denoted as CCL) is compared against the one 
based on the original data with the complete rankings regarded here as a ground truth and the one 
based on the marginal probabilities treated as independent (denoted as ICL). The ICL accuracy on the queries is evaluated separately on the pairs on which CCL is determinate and the ones on which CCL is indeterminate (i.e., indecision is returned). These results are in Figure~\ref{fig:exps}. The separation between these two accuracies is clear: 
 CCL becomes undecided on the tasks on which a less conservative approach would be less accurate, thus providing a more robust approach to the inferences.

\section{Conclusions}\label{sec:conc}
In the last years, several works in the probabilistic logic tradition have proposed formalisms  to explicitly deal with independence, see e.g. \cite{andersen1994bayesian,cozman2008probabilistic,haenni2010probabilistic,michels2015new}. Our approach differs in the sense that,  rather than attempting to combine probabilistic logics and probabilistic networks, is closer to logic programming. Indeed, we have introduced CCL, a conservative extension of Poole's ICL in which both independent and non independent choices can be modelled. In the proposed setting, a theory specifies the (credal) set of all probability mass functions over least models compatible with the marginals  on atomic choices. In guise of example, we applied CCL to object ranking and have shown how to  infer the lower and upper probabilities of a query. 
In future work, on top of presenting more complex applications and deeper experiments of the proposed formalism, we plan to compare our work with related  approaches such as for instance the one discussed by \cite{flesca2014consistency} in the context of probabilistic databases. 
\bibliographystyle{splncs04}
\bibliography{biblio}
\end{document}